# INFLATABLE DEVICES FOR PLANETARY AEROCAPTURE AND AEROBREAKING MANOEUVRES

Philippe Reynier[1]

*Ingénierie et Systèmes Avancés, Cestas, France*

Philippe.Reynier@isa-space.eu

Future missions to Mars and Venus will make use of aerobraking and aerocapture in order to gain mass through the saving of fuel at planetary arrival. So far only aerobraking has been demonstrated, if the Mars Premier project has paved the way for aerocapture, no demonstration was performed due to the project interruption. The use of these techniques induces additional constraints for planetary probes, since additional heating and mechanical loads have to be carefully managed. Moreover, aerocapture requires a high level of accuracy for the Guidance Navigation and Control aspects, since a pass at an altitude of the atmosphere with a different density could lead to the vehicle destruction. This document surveys the existing state-of-the-art on inflatable devices (including ballutes, sails, or inflatable heat-shield capsule) for orbital manoeuvres at planetary arrival.

## Nomenclature

**Acronyms:**

CFD: Computational Fluid Dynamics;
CNES: Centre National d'Etudes Spatiales;
ESA: European Space Agency;
GNC: Guidance, Navigation, and Control;
IRDT: Inflatable Re-entry Demonstrator Technology;
IRVE: Inflatable Re-entry Vehicle Experiment;
NASA: National Aeronautics and Space Administration;
TPS: Thermal Protection System.

[1] Research Engineer, ISA, 33610 Cestas, France, Philippe.Reynier@isa-space.eu

## 1 Introduction

This survey focuses on the investigation of inflatable devices to be used for aerobraking and/or aerocapture of exploration probes to planets. Aerobraking [1] consists of performing a cycle of orbits from an elliptic transfer orbit to the final one as shown in Figure 1, at high altitude, in order to decrease the spacecraft velocity. At each loop the spacecraft crosses the upper layer of the atmosphere, the flow around the vehicle remains rarefied avoiding the apparition of a bow shock and subsequent aeroheating that could not be sustained by the spacecraft.

Aerocapture can be performed only at spacecraft arrival, from a hyperbolic trajectory, to the planet; in order to save some propellant mass one pass is performed in the atmosphere, but not as for aerobraking in the upper layer, but at lower altitude. As a consequence, this manoeuvre requires a Thermal Protection System (TPS) for the spacecraft. At the end of the pass, the spacecraft is inserted on its final orbit. Both aerobraking and aerocapture are attractive since they allow large savings in terms of propellant needed by the spacecraft as highlighted for different mission in [11], however, these techniques are very challenging since they require high level of fidelity in the GNC subsystem, and an accurate knowledge of the planet atmosphere that can present some density variability. First aerobraking was performed around Earth by the Japanese probe Hiten in 1991, the European Space Agency (ESA) performed is first with Venus Express in 2006, so far, no aerocapture has been performed, however, this option was retained for the Mars Premier project developed by CNES and NASA and interrupted in 2003. Since, other studies have been done for missions to Mars, Titan, and Ice Giants [2-3] involving aerocapture but so far none was performed.

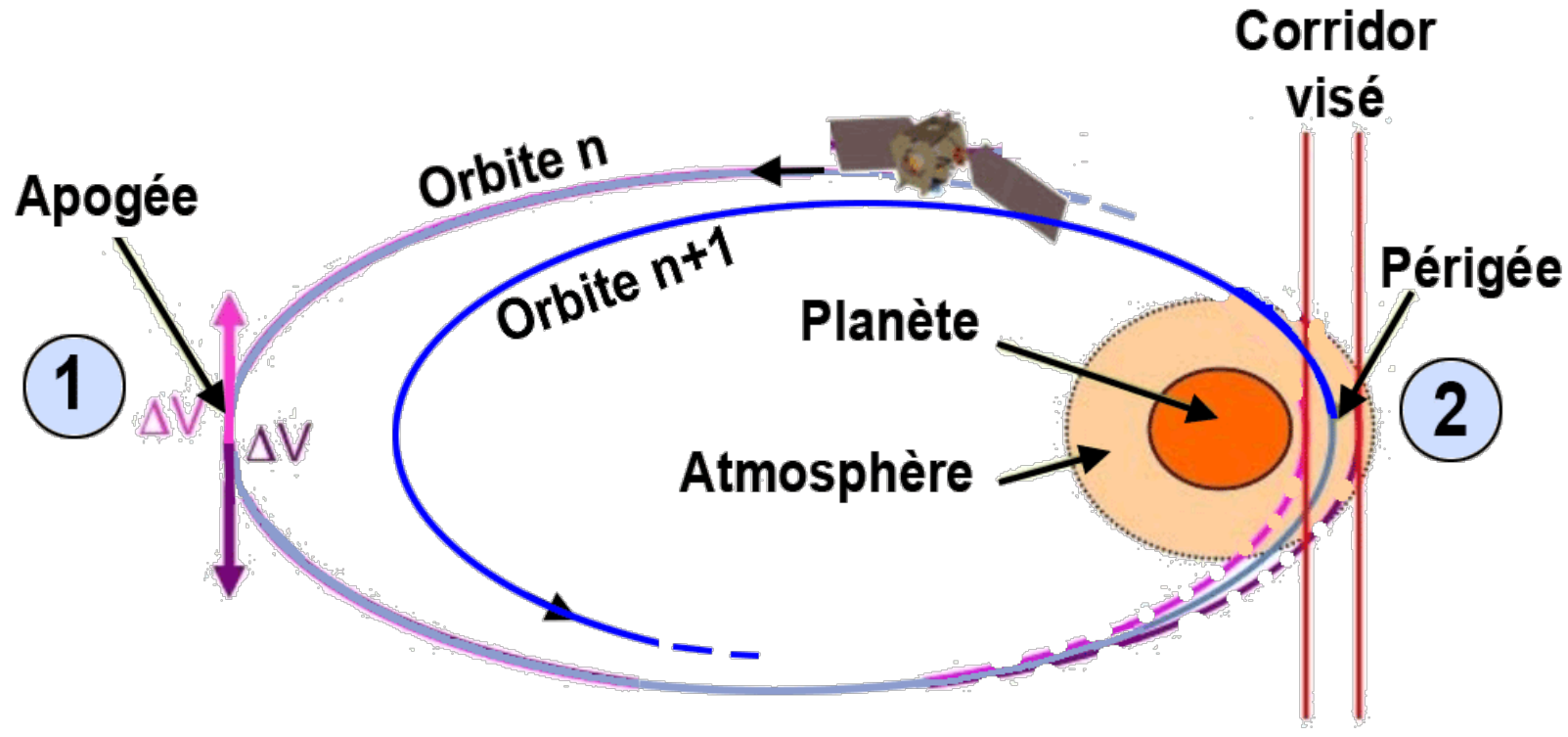

**Figure 1: Sketch of aerobraking with targeted corridor (source Wikipédia)**

## 2 Scenarios & Configuration

A large number of studies can be found in the literature focusing on inflatable devices for performing atmospheric entry, aerobraking, and aerocapture. Concerning entries involving inflatable devices, the pioneering flights tested technology was the series of IRDT (Inflatable Re-entry Demonstrator Technology) [4-5], shown in Figure 2, developed by the Babakin Space Center in Russia in the frame of an ESA R&D programme in the early 2000's. As far as aerobraking and aerocapture are concerned, most of the literature on potential inflatable devices focuses on ballutes, little elements can be found on sails or deployable devices that could be relevant for aerobreaking and aerocapture respectively.

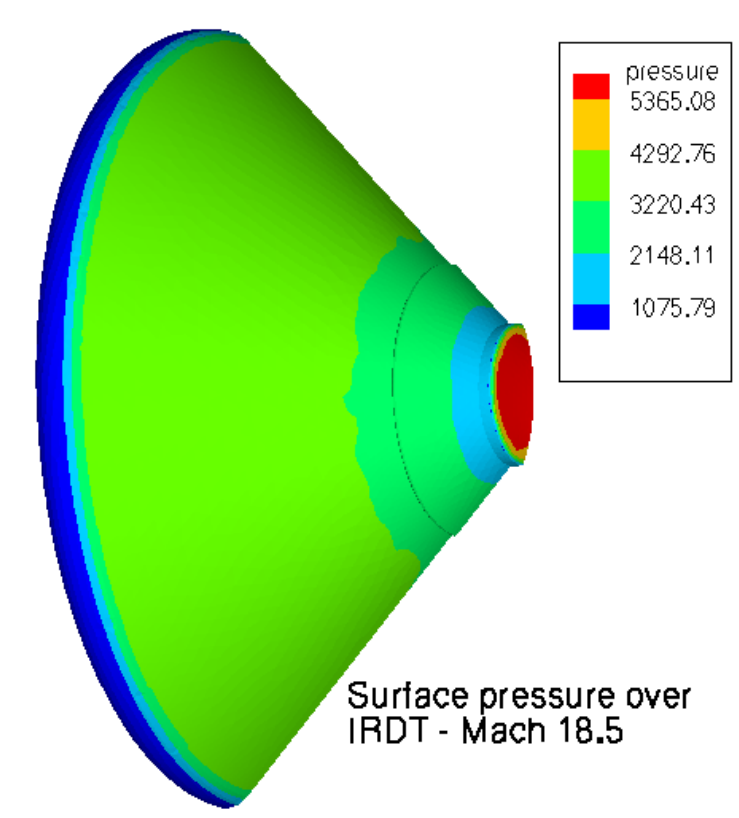

**Figure 2: Surface pressure over IRDT [4]**

The ballute concept has been studied for aerocapture to Venus, Mars, Neptune, Titan, and Saturn [6-7-8-9]. As a consequence, the corresponding existing state-of-the-art on ballutes is here surveyed but the elements available on deployable sails for such missions are also considered since a sail is not so different from a ballute, from the breaking device aspect.

For aerobreaking and aerocapture, the spacecraft hyperbolic velocity at planet arrival as well as the spacecraft cross-section area, are two key parameters. Their knowledge is a necessity for estimating the drag coefficient, and therefore the energy dissipated, and the mechanical loads, during aerobraking loops (or aerocapture pass). Concerning the aerothermodynamics and aerodynamics, aerobraking (and most of an aerocapture) is expected to occur in a rarefied regime, therefore the aerothermal loads will remain low, and drag and mechanical loads can be estimated using existing correlations (for example stagnation point correlations for aeroheating). In addition, solar heating will have to be considered since it will not be negligible when comparing to the aerothermal heating during a Venus aerobraking [10].

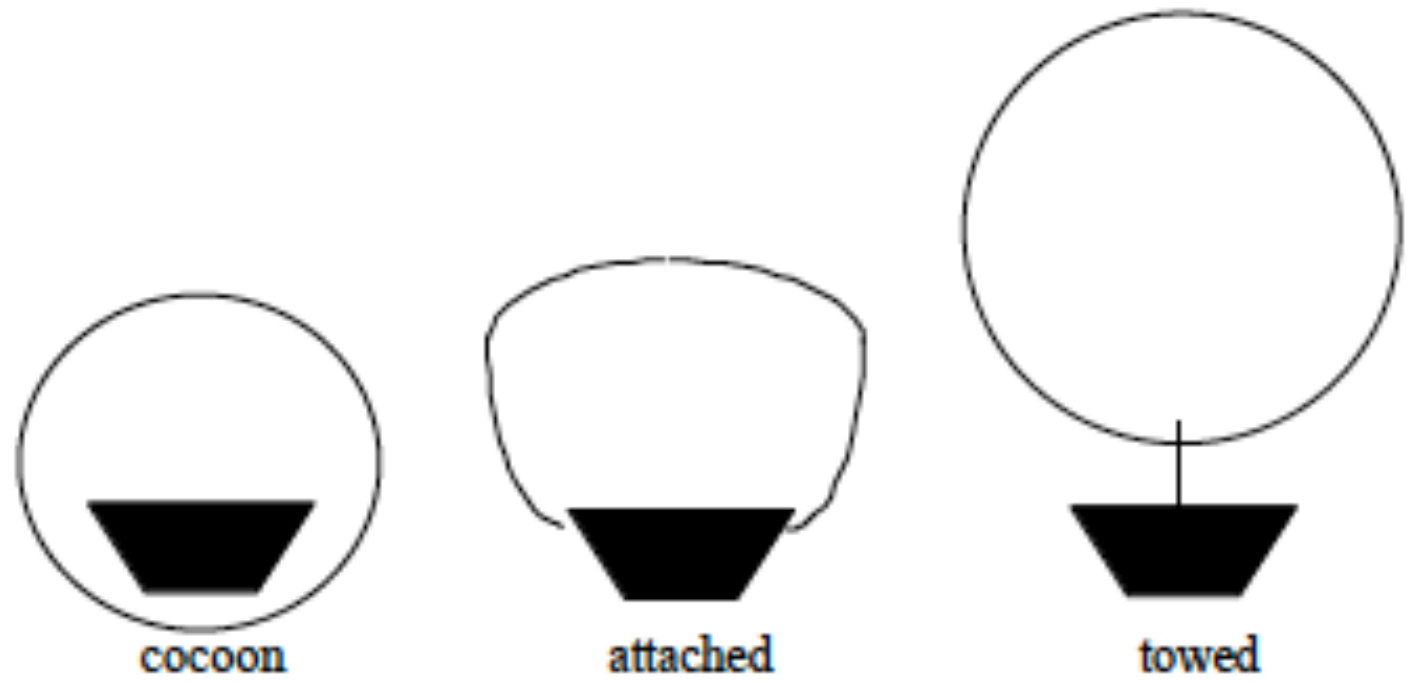

**Figure 3: The three different ballute configuration [11]**

Another point related to the configuration is the spacecraft shape, important to have the cross-section area for evaluating the drag, but also to size the inflatable device and assess the potential interactions with the spacecraft. Finally, this is also an important issue for the spacecraft attitude and stability; this last point is crucial if an aerocapture has to be performed.

The first point to be investigated is the configuration of the inflatable device, it can be attached, cocoon, or towed as shown in Figure 3, or part of the thermal protection system as highlighted in Figure 2. According to Hall [11], for a ballute towed is the preferred configuration. It avoids the cocoon problem of having to deploy and then remove a membrane around fragile spacecraft components like solar panels, antennae, and scientific instruments. Second, it physically separates the ballute (or the sail if this device is used) and the spacecraft, thereby preventing lateral ballute forces from affecting the orientation of the spacecraft. This can be particularly important given the likelihood for unsteady shock waves and vortex shedding associated with the flow around ballutes. Third, a towed ballute is easily detached from the parent spacecraft by cutting the connecting tether. Given the large drag disparity between the ballute and spacecraft, this enables the drag to be essentially "turned off" once the desired decrease in vehicle kinetic energy has been achieved. Ballutes can be of different shapes: spheres, disks, or toroid's (see Figure 4) are generally retained.

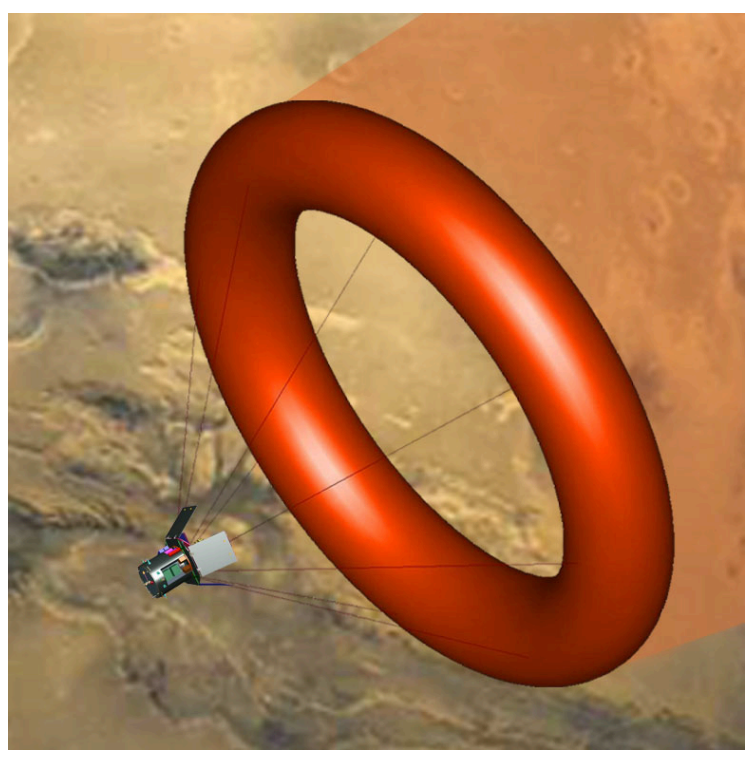

**Figure 4: Toroidal ballute for aerocapture [7]**

Comparing to a ballute, a deployable sail will have the advantage to present higher manoeuvrability qualities, but might be very fragile (depending on the blanket thickness) to heating and to particle impacts [12]. This last point could be a concern for a Mars mission due to the potential presence of dust at high altitude. These different points will be assessed in the following section and the related trade-offs identified.

## 3 System trade-offs and issues

The main objective of this section is to review the existing state-of-the-art for inflatable devices in order to assess their advantages and potential drawbacks. A certain literature is already available [3,13] on some points such as deployable booms, storage, inflatable material aspects, and tethers. The problem areas presenting technical issues are also identified in this section. The identified trade-offs related to the different inflatable devices, in the perspective of aerobreaking or aerocapture, are:

- Thermal and structural issues;
- Aerocapture efficiency;
- Aerobraking efficiency;
- Dimensions and storage;
- Mass versus deployed area;
- Pressurised device (ballute, inflatable capsule IRDT-like) or deployable (sail);
- Area versus dynamic behaviour.

In addition, the following elements [6] have to be considered for a potential device selection:

- A device much larger than the parent spacecraft drastically shifts the trajectory to higher altitudes where the density is very low.

- Large structures in low-density flows experience much reduced heating to such a point that flexible, balloon-like materials can survive.
- Low device masses can be achieved by using thin-film material that does not ablate but self cools by thermal radiation.
- The device flies a ballistic trajectory in which the only modulation is timing the moment of ballute detachment from the spacecraft.
- The device is towed behind the spacecraft to facilitate an easy detachment and to prevent possible interferences with spacecraft functions like telecommunications and thermal management.

The operational scenario: aerocapture, aerobreaking, characteristics of planet atmosphere; proposed has to be critically evaluated to evaluate the corresponding operational envelopes and constraints in terms of thermal and mechanical loads, and system and design impact.

The trade-offs to be performed for selecting an inflatable device have to be mostly based on literature data' this is due to the lack of experimental datasets. They shall include system analysis, experimental data, and numerical calculations performed for inflatable devices. The state-of-the-art supporting this study shows that if ballutes are considered, towed ones are more attractive than the other options (cocoon or attached). The most promising device for an aerobreaking is a towed one the final trade-off is between a ballute (most probably a toroïdal one) and a sail (most probably in a pyramid shape). However, not depending on this trade-off several potential problems can be already identified [7]:

- Packing and Storage – The years of transit for some of the outer planet missions put a requirement on the device to work after being stored for a very long time. There is limited data on the properties of some of the candidate materials after long-term storage. Blocking, creasing and pinholes due to packing may present problems for the inflatable as well.

- Deployment – Controlled deployment and inflation are critical (this particularly applies for long missions for which it occurs after several years of storage in space conditions). To mitigate any risk of tethers getting tangled and to avoid re-contact between the spacecraft and the device, the baseline design concept could include inflatable columns in some tether locations to provide some compressive stiffness in the system.

- Aeroelasticity – Interactions between the hypersonic, rarefied flow and the ballute could present a problem. This is very complex to assess analytically, and facility limitations make testing very difficult. It can be argued that the rates between the very high speed, low density flow and the lower resonance of the inflatable make aeroelasticity at the altitudes that are being considered for the thin-film ballute concept incompatible; however, verification of this is challenging. This applies also for inflatable devices if an aerocapture is foreseen, the tearing of the inflatable material during the inflation has to be avoided for ensuring the integrity of the spacecraft (this issue was responsible of the failure of the last IRDT test in 2005).

- Material – Thermal, aerial density, tensile strength, material maturity, ballute manufacturability, and seaming characteristics are all critical factors in the design. Candidate materials such as Kapton or similar have to be evaluated. For aerocapture, an ablative blanket might have to be used as thermal protection system for the spacecraft, the aerothermal loads are to high for materials like krypton and low enough to avoid a rigid material.

- Flow Stability – The potential exists for the toroidal concept of choked flow through the torus. The subsonic flow in the wake of the spacecraft provides a means of back flow that can disrupt the flow stability for any ballute or sail devices.

| Issues | Description |
|---|---|
| Optimal shape | Which shape (sphere, disk, toroid, sail) for which missions? |
| Survivability | Can the membrane of the TPS (for aerocapture) survive the heating and the mechanical loads? |
| Stability | If the flow is unstable, can this be sustained? |
| Trajectory | How to compensate the atmospheric uncertainties? This is a key point for aerocapture. |
| Structural aspects | How to ensure the device integrity? |
| Device design | Can the device be designed for handling heating and mechanical stress? |
| Parent spacecraft protection | What does specific protection (for aerothermal effects) the spacecraft require? What is the impact on the mass budget? |
| Deployment and inflation | Is there an existing heritage available? Or an available facility for verification? |

**Table 1: Potential issues for inflatable devices**

Some potential issues for ballutes have been already identified by Hall [11]; they apply for the different inflatable devices and are listed in Table 1. One of the more crucial points is the stability of the device that has to be maintained during the inflation and the manoeuvre. This can be performed through computational simulations as done in [14] for the stability of solar sails when mounted on-board the spacecraft for performing end-of-life re-entry.

Another point concerns the aerodynamics around both spacecraft and inflatable device. If for an aerobraking the flow will remain rarefied, this might not be the case for the whole aerocapture. A particular point in the possible presence of continuum flow around the spacecraft while the device would have to sustain transitional or rarefied flow conditions. Moreover there is the possible problem of significant plasma radiation (depending on the planet atmosphere and vehicle velocity) that would require materials with a good level of opacity in the relevant wavelength ranges.

## 4 Current state-of-the-art

In the following a survey of the state-of-the art on inflatable/deployable devices for aerobraking and aerocapture is performed. Most of the literature material used for this survey has been obtained for ballutes.

Most of the European heritage on aerocapture studies has been performed by CNES for Mars Premier [15-16] and at ESA in the frame of the Mars Aerocapture Group [17-18]. The same remark applies for inflatable systems with the ESA activities carried out for the development of for atmospheric re-entry demonstrator IRDT [4].

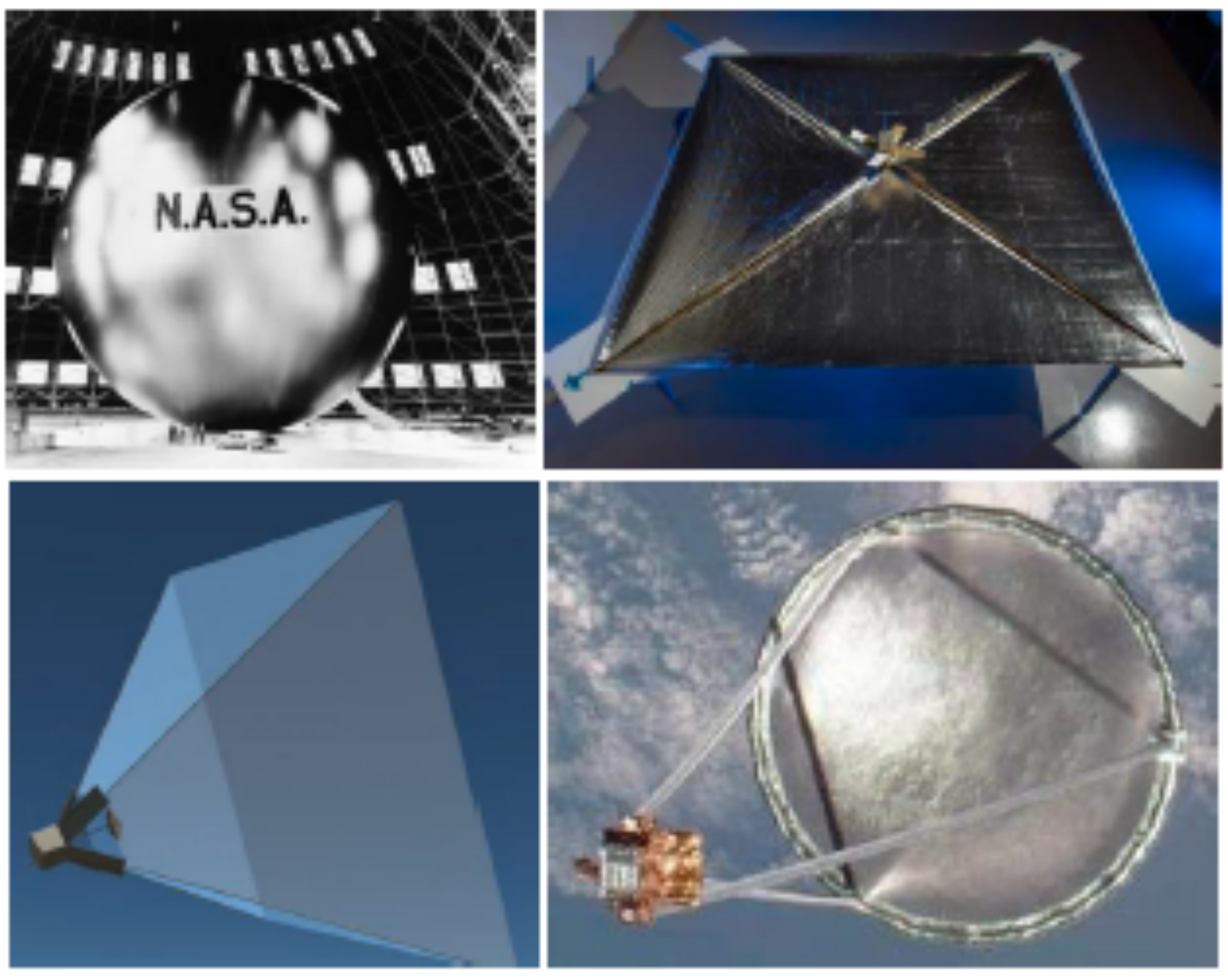

**Figure 5: Typical deployable sail shapes [21]**

In the frame of Venus Express mission, first European aerobraking was performed. Different elements on GNC, mission design, trajectory and models are available in [10]. Aerodynamics as well as heating were assessed. Solar heating was found to be not negligible when comparing against aerothermal loads. Main issues for Venus aerobraking were found to be the atmosphere variability and telemetry. For Mars, aerobraking was studied for the Mars Express mission and performed for ExoMars. Different contributions [19-20] focused on the flight dynamics aspects. The vehicle was covered by a thermal multi-layer insulation, rarefied dynamic heating was considered for the design and analysis. The corridor was designed [20] to have a dynamic pressure lower than 0.28 Pa, a heating below 1.1 kW/m$^2$ and a heat-load lower than 200 kJ/m$^2$. The problem of variability of Mars atmosphere was also considered to be an issue.

Another attractive device for aerobraking is to use a deployable sail. This device has been already studied as de-orbiting device [12]. Typical shapes of deployable sails are shown in Figure 5. However, the study mentioned is mostly focused on GNC. Specific issues associated to this device are the sail degradation due to micrometeoroid impacts, and the importance of solar heating.

## 4.1 Ballutes

### 4.1.1 *Aerodynamics & Design*

According to Gnoffo & Anderson [8], a toroidal system can be designed (ignoring influence of the tethers) such that all flow processed by the bow shock of the towing spacecraft passes through the hole in the toroid. For a spherical ballute, towline length is a critical parameter that affects aeroheating on the ballute being towed through the spacecraft wake. In both cases, complex and often unsteady interactions ensue in which the spacecraft and its wake resemble an aerospike situated in front of the ballute. The strength of the interactions depends on the system geometry and Reynolds number. The same authors show how interactions may envelope the base of the towing spacecraft or impinge on the ballute surface with adverse consequences to its thermal protection system. Geometric constraints to minimize or eliminate such adverse interactions are discussed. The towed, toroidal system and the clamped, spherical system show greatest potential for a baseline design approach.

Finally, a very positive variation of the towed spherical ballute system employs a zero towline length (clamped system). In this design the spacecraft is contained fully within the low subsonic shock layer of the ballute and is nearly engulfed within the ballute boundary layer. Heating to the spacecraft forebody is reduced by a factor of two or more (for the particular system considered here) and no adverse aeroheating is induced on the ballute [8]. Towlines and associated deployment systems are obviated in this design concept.

| Temperature (C) | Kapton film | PBO film (machine direction) | General PBO film | High performance PBO fibre |
|---|---|---|---|---|
| 20 | 21.1 | 98.4 | 56-63 | 576 |
| 100 | 15.8 | 84.4 | | |
| 200 | 10.8 | 63.3 | | |
| 300 | 7.7 | 49.2 | | |
| 400 | 5.55 | 45.7 | | |
| 500 | 3.94 | 30.9 | | |

**Table 2: Kapton and PBO film strength (kg/mm$^2$) as function of temperature [13]**

#### 4.1.2 *Material aspects*

The survivability of the ballute material is one of the key issues of the system. However, the use of very large ballute diameter decreases drastically the heating since this last varies as the inverse of the square root of the radius of the curvature.

A central feature of the towed ballute concept is that the membrane material will self-cool by radiated energy emission at a tolerably low temperature [11]. Literature data indicate that temperatures up to 770 K can be sustained with Kapton thin films. Tensile strength of Kapton is 38 MPa, significant but not outstanding. Another material of interest is polybenzoxazole (PBO) either in thin film or fabric form.

For a typical mission considering a ballute, this last has to be capable of surviving an aeropass (remaining inflated) for approximately ten minutes. According to Gnoffo & Anderson [8] laminates of Kapton film with polybenzoxazole (PBO, a liquid crystal polymer) are believed to combine the needed tensile strength and temperature limits (up to 770 K) required in this application. Tensile strength as function of temperature is reported in Table 2. For tensile tethers, high strength materials such as Kapton and Twaron [7] are available, more advanced materials are under development. Specific adhesives have to be selected for bonds and coatings; these points are extensively discussed in [7].

The mean free path of the free-stream during the release of a typical ballute is about 1 m, so that the flow then is approximately continuum, while it will be free-molecular on the tubing and intermediate on the orbiter. It is normally accepted that a Knudsen number of 0.01 or lower is necessary for having a fully developed bow shock, and in the aerocapture case the ballute is probably released before this value is reached. In the entry case however, much of the drag occurs in developed continuum conditions, and a ring shaped ballute is indicated.

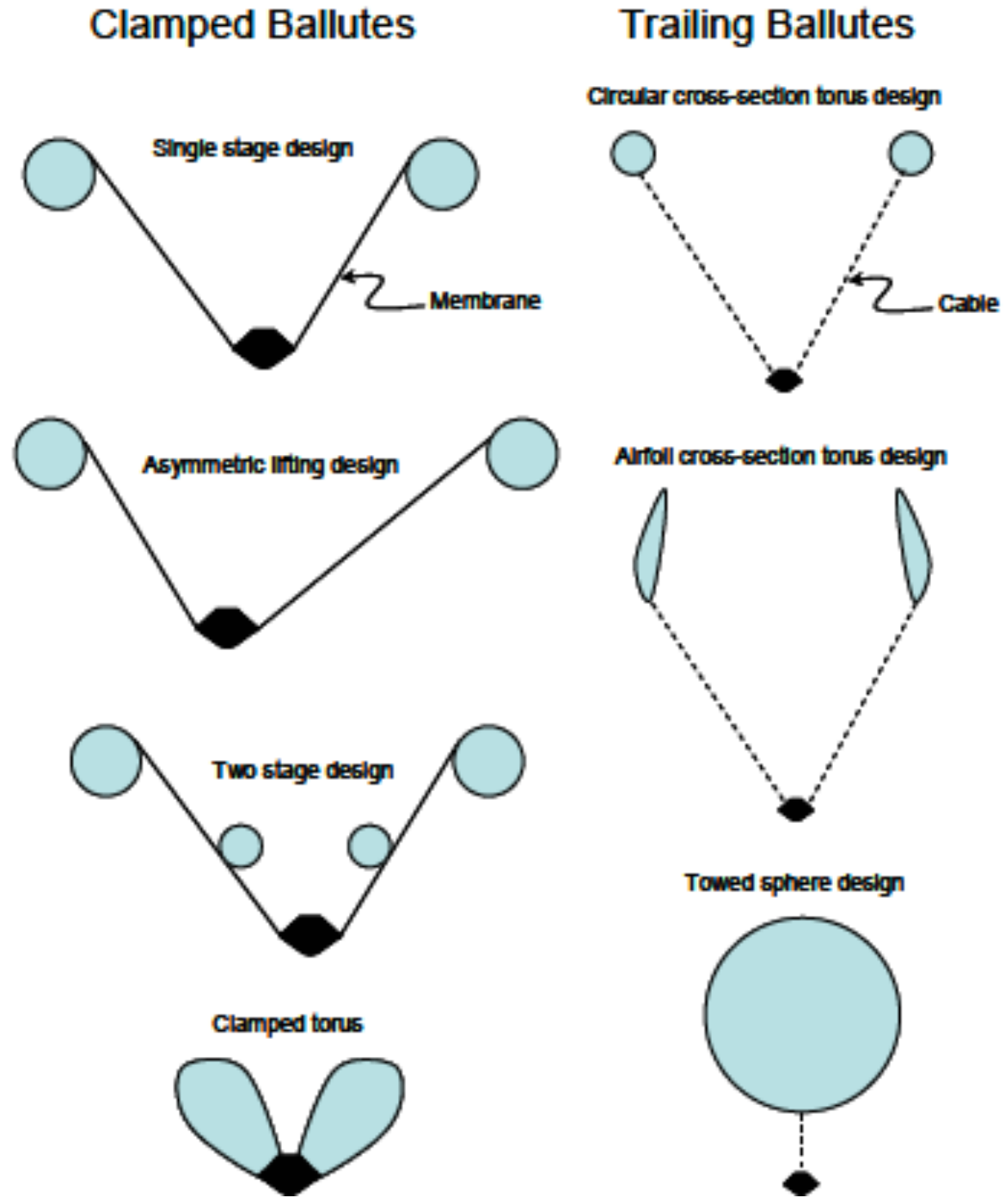

**Figure 6: Examples of clamped and trailed ballutes [21]**

4.1.3 *System aspects*

There are three different concepts of ballutes: cocooned, attached and towed as depicted in Figure 3, they can be also of different shapes [21], symmetric or not as shown in Figure 6. According to Hall [11], the towed ballute is currently viewed as the preferred configuration for planetary aerocapture because of three key advantages. "*First, it avoids the cocoon problem of having to deploy and then remove a membrane around fragile spacecraft components like solar panels, antennae, and scientific instruments. Second, it physically separates the ballute and spacecraft, thereby preventing lateral ballute forces from affecting the orientation of the spacecraft. This can be particularly important given the likelihood for unsteady shock waves and vortex shedding associated with the flow around ballutes. Third, a towed ballute is easily detached from the parent spacecraft by cutting the connecting tether. Given the large drag disparity between the ballute and spacecraft, this enables the drag to be essentially "turned off" once the desired decrease in vehicle kinetic energy has been achieved*".

The inflation pressure can be made high enough to provide appreciable stiffness to the ballute membrane, thereby enabling non-parachute shapes like cones, toroids and sphere.

A key premise of the concept is that the timing of ballute detachment provides sufficient trajectory modulation capability to enable aerocapture despite atmospheric uncertainties and navigation errors at the target planet. Numerical simulations [6] performed for four candidate (see Table 3) missions have confirmed this, and quantified the entry corridor in terms of initial flight path angle and zero-drag periapsis altitude ranges at each planet. Corridors exist for large, towed aerocapture ballutes using only the time of ballute detachment as a modulation variable. It appears that the computed corridor widths are sufficient to accommodate the anticipated approach navigation and atmospheric density profile errors for these missions. A shared result across these missions is that the zero-drag periapsis altitude range divided by the density scale height of the atmosphere yields, a common value of approximately 2 despite a wide range of entry velocities, spacecraft masses and atmospheric compositions.

| Mission | Venus Sample Return | Mars Microsat | Saturn Ring Observer | Titan Organics Explorer | Neptune Orbiter |
|---|---|---|---|---|---|
| Atmosphere, $\chi_{i,\infty}$ | $CO_2$ = .965<br>$N_2$ = .035 | $CO_2$ = .953<br>$N_2$ = .027<br>Ar = .016<br>$O_2$ = .004 | $H_2$ = .963<br>He = .037 | $N_2$ = .983<br>$CH_4$ = .017 | $H_2$ = .80<br>He = .19<br>$CH_4$ = .001 |
| Sphere, D, m | 135. | 12. | 120. | 48. | 100. |
| $V_\infty$, km/s | 10.6 | 5.54 | 24.1 | 8.53 | 26.7 |
| $\rho_\infty$, kg/m$^3$ | 2.2 $10^{-7}$ | 1. $10^{-6}$ | 3.1 $10^{-8}$ | 3.0 $10^{-7}$ | 1.5 $10^{-8}$ |
| $C_D$ | 0.913 | 0.99 | 0.98 | 0.981 | 1.05 |
| $\dot{q}_{max}$, W/cm$^2$ | 1.07 | 1.82 | 2.72 | 1.93 | 2.60 |
| $T_{max}$, K | 677. | 774. | 855. | 785. | 845. |
| Toroid, m<br>Ring D / X-sec D | 150. / 30. | 15. / 3. | 120. / 30. | 52. / 13. | 100. / 25. |
| $V_\infty$, km/s | 10.6 | 5.49 | 23.9 | 8.55 | 26.8 |
| $\rho_\infty$, kg/m$^3$ | 1.6 $10^{-7}$ | 7.1 $10^{-7}$ | 2.3 $10^{-8}$ | 1.9 $10^{-7}$ | 8.2 $10^{-9}$ |
| $C_D$ | 1.31 | 1.45 | 1.38 | 1.39 | 1.51 |
| $\dot{q}_{max}$, W/cm$^2$ | 1.31 | 1.91 | 2.85 | 2.05 | 2.84 |
| $T_{max}$, K | 712. | 783. | 865. | 796. | 864. |

**Table 3: Ballute simulations for different planetary entries [9]**

During an aerocapture, if too much energy is removed before leaving the atmosphere (see Figure 7), a drag makeup manoeuvre [7] might be required to raise apoapsis out of the atmosphere. A manoeuvre at the first apoapsis is essential for raising periapsis out of the atmosphere. Another important shared feature is that the aerocapture heating pulse is of short duration with a maximum value at approximately 250 seconds for the most challenging missions.

To summarize, these are the essential elements of the large, towed aerocapture ballute concept:

- A ballute much larger than the parent spacecraft drastically shifts the trajectory to higher altitudes where the density is very low;
- Large structures in low density flows experience much reduced heating to the point that flexible, balloon-like materials can survive;
- Low ballute masses can be achieved by using thin-film material that does not ablate but self cools by thermal radiation;
- The ballute flies a ballistic trajectory in which the only modulation is timing the moment of ballute detachment from the spacecraft;
- The ballute is towed behind the spacecraft to both facilitate easy detachment and to prevent interference with spacecraft functions like telecommunications and thermal management.

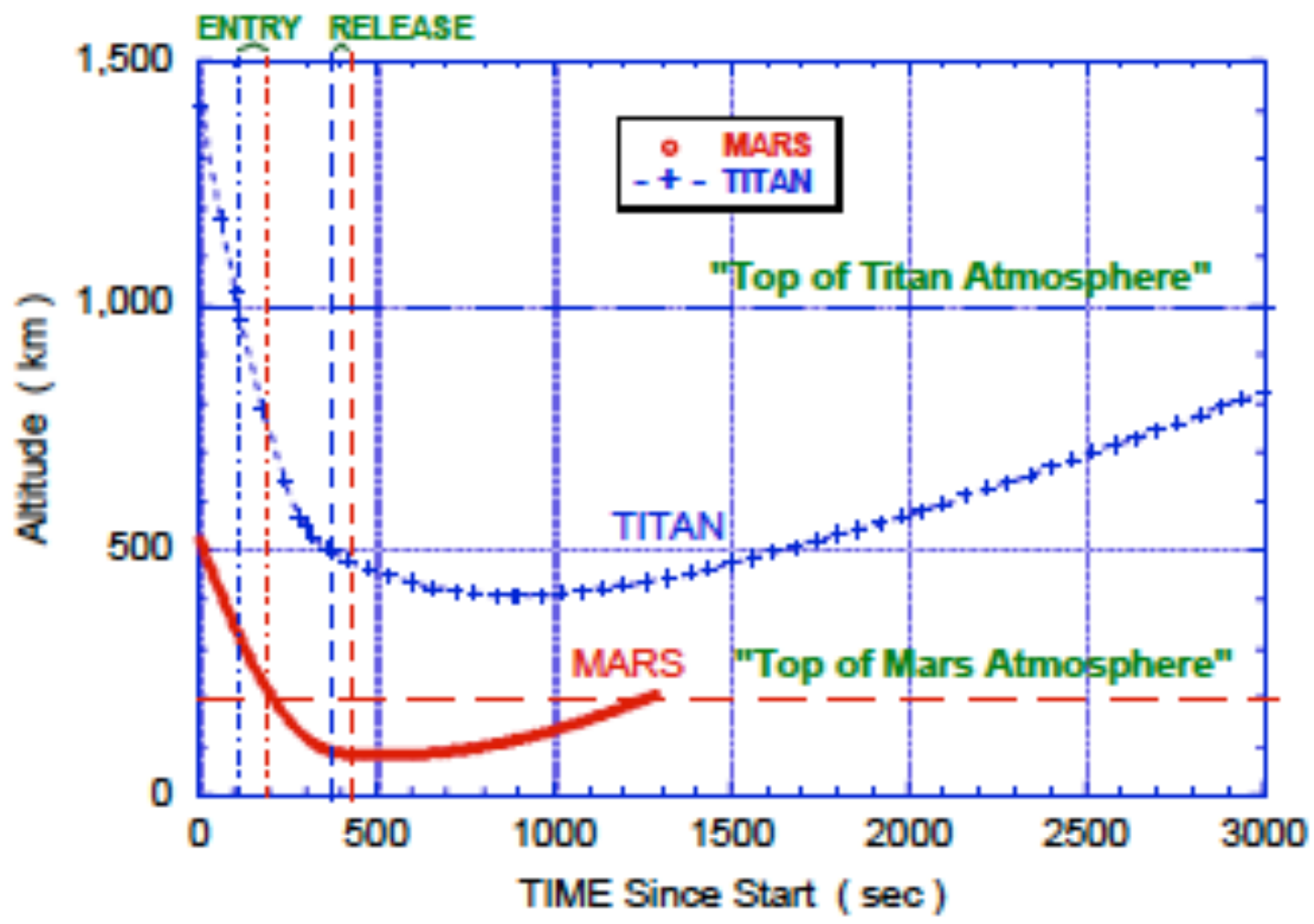

**Figure 7: Altitude versus time for aerocaptures with ballute at Mars and Titan [7]**

The different issues are resumed in Table 1, they have been extensively discussed by Miller et al [7], a spherical ballute is the simplest towed geometry, one, according to Hall [11], that possesses two key advantages: the drag and heating is insensitive to ballute orientation and the ballute requires only a single tether whose own heat load will be minimized due to its alignment with the velocity vector. The main disadvantages of the sphere are that the envelope mass and the mass of the inflation gas, grow quickly with the ballute diameter. Calculations suggest that these masses may become prohibitive ballutes required for the Venus and Saturn missions reported in Table 3. Alternatives to spheres include a disk and the toroidal ballute that can be more or less massive than the sphere depending on the ratio of ring to cross section diameters. However, both the disk and toroidal ballutes are orientation sensitive and hence they require multiple tethers that may see appreciable heating due to only partial alignment with the flow.

For an aerocapture the ballute has to be deployed before entering into the atmosphere, for such a purposed, the technology can be borrowed to current state-of-the-art. An important point is the gas pressure inside the ballute that will be increased due to the heating during the aerocapture pass. A mechanism for the pressure monitoring will have to be integrated with for consequence a penalty in term of mass.

There are a number of different facets to the issue of flow stability of the towed ballute. First, there is the problem of large lateral forces on the ballute due to vortex shedding that can alter the orientation of the ballute and the parent spacecraft. Second, there can be aeroelastic phenomenon associated with the non-rigid nature of the ballute leading to possible shape changes under aerodynamic loads. Third, there is the possibility of flow instability due to wake and shock wave interactions between the ballute and spacecraft.

#### 4.1.4 *Aerothermal studies*

In order to support the aeroassisted entry using ballutes several studies have been focused on the numerical and experimental investigation of aerothermal loads surrounding these spacecraft [8-9]. The assessment can be supported by available experimental and numerical results from the literature. Several test campaigns have been conducted in shock-tunnels [22] and expansion tube [23] for investigating heating and flow-field topology. Test results, depending on flow conditions have put in evidence interaction between the wake of the spacecraft and the inflatable device as highlighted in Figure 8, as well as the presence of flow unsteadiness as shown in Figure 9, both phenomena can affect the device stability. In addition, numerical simulations [8-9] carried out for evaluating shock-layer and wake behaviour, as well as unsteady interactions are necessary to ensure the stability of the device. The spacecraft wake impact, is very similar for a sail, if we consider the possible fluid/structure interaction with the sail sides.

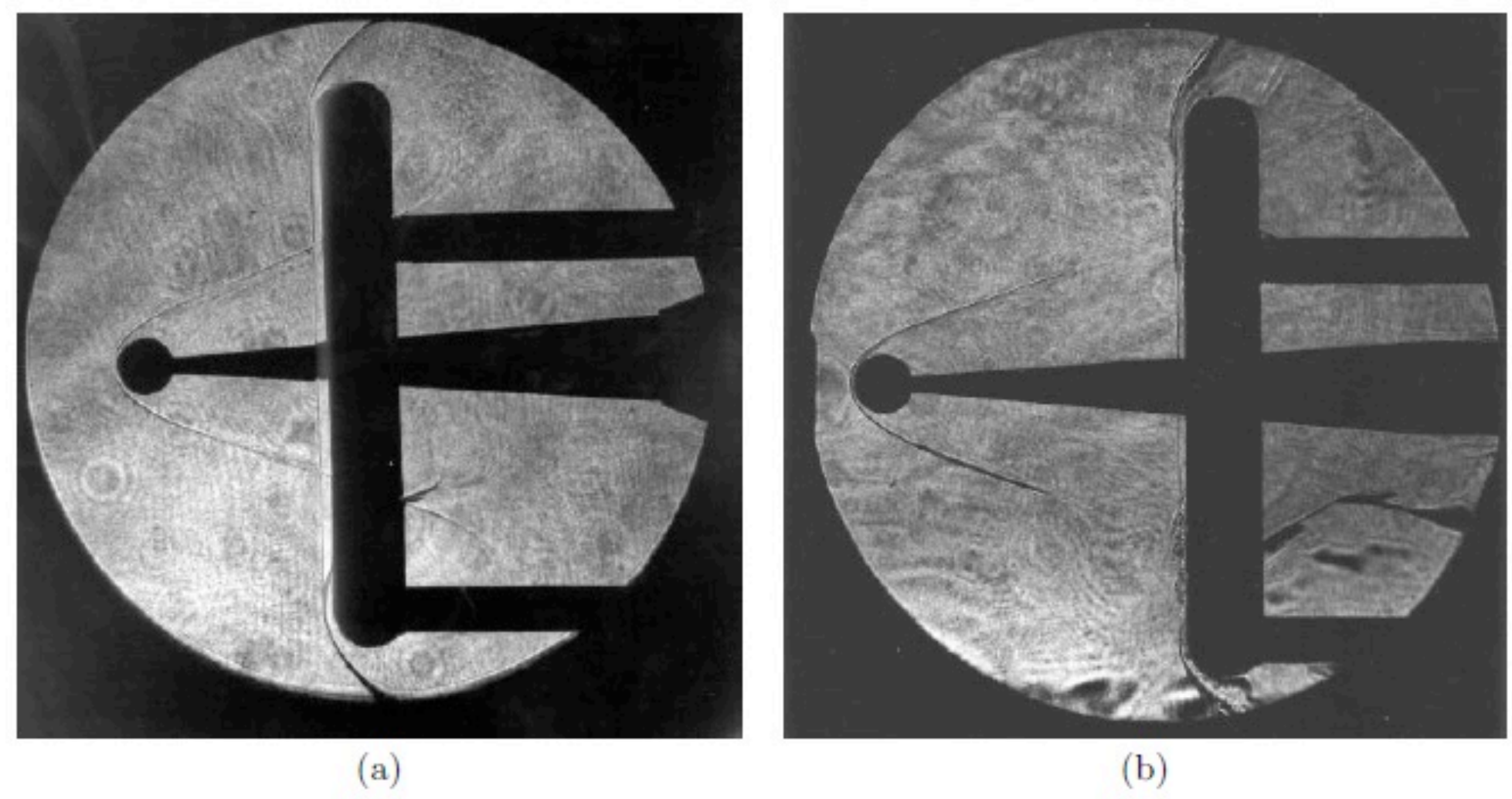

**Figure 8: Shadowgraphs for a $CO_2$ flow at 17.7 and 20.4 MJ/kg in T5 [22]**

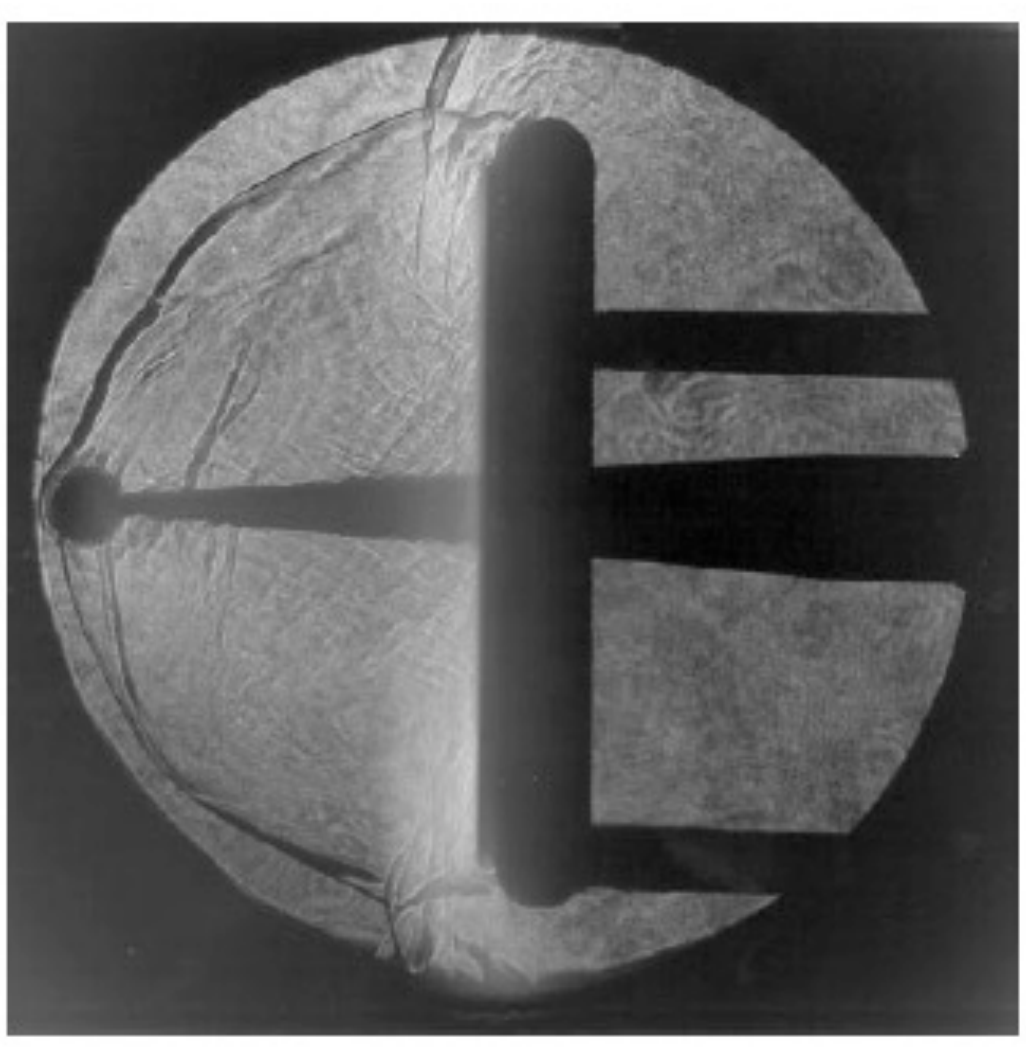

**Figure 9: Presence of flow unsteadiness in $N_2$ at 25.6 MJ/kg [22]**

In [9], a towed toroid configuration is first computed, then the MSRO [24] aerocapture is investigated. The different simulations are resumed in Table 3. The toroid configuration has the advantage to have the spacecraft wake passing within the toroid hole. The vibrational temperature contours for Mars conditions are shown in Figure 10. They highlight the presence of a high temperature core behind the shock-shock interaction in the near wake. Here, shock stand-off of several meters is computed. Experimental campaign [22] performed in T5 shock-tunnel with a stainless-steel toroid ($V_\infty$= 3714 m/s, $CO_2$, $\rho_\infty$ = 1.61 kg/m$^{-3}$, $T_\infty$ = 1570 K) was also numerically reconstructed. Comparisons against aeroheating data are plotted in Figure 11. They show heating levels bounded by two strongly catalytic models for recombination at the surface, with however a fair agreement.

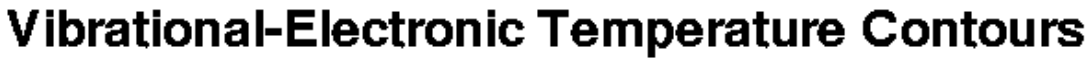

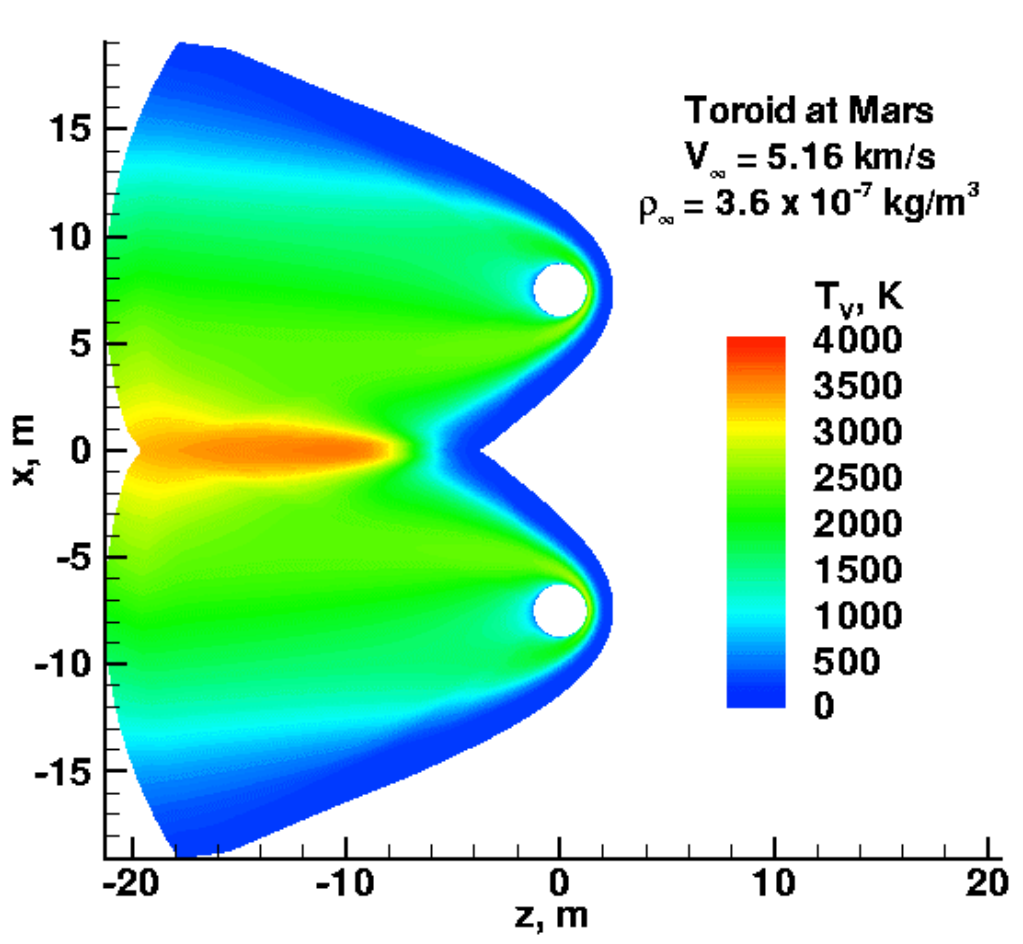

**Figure 10: Vibrational temperature around a toroid for Mars conditions [9]**

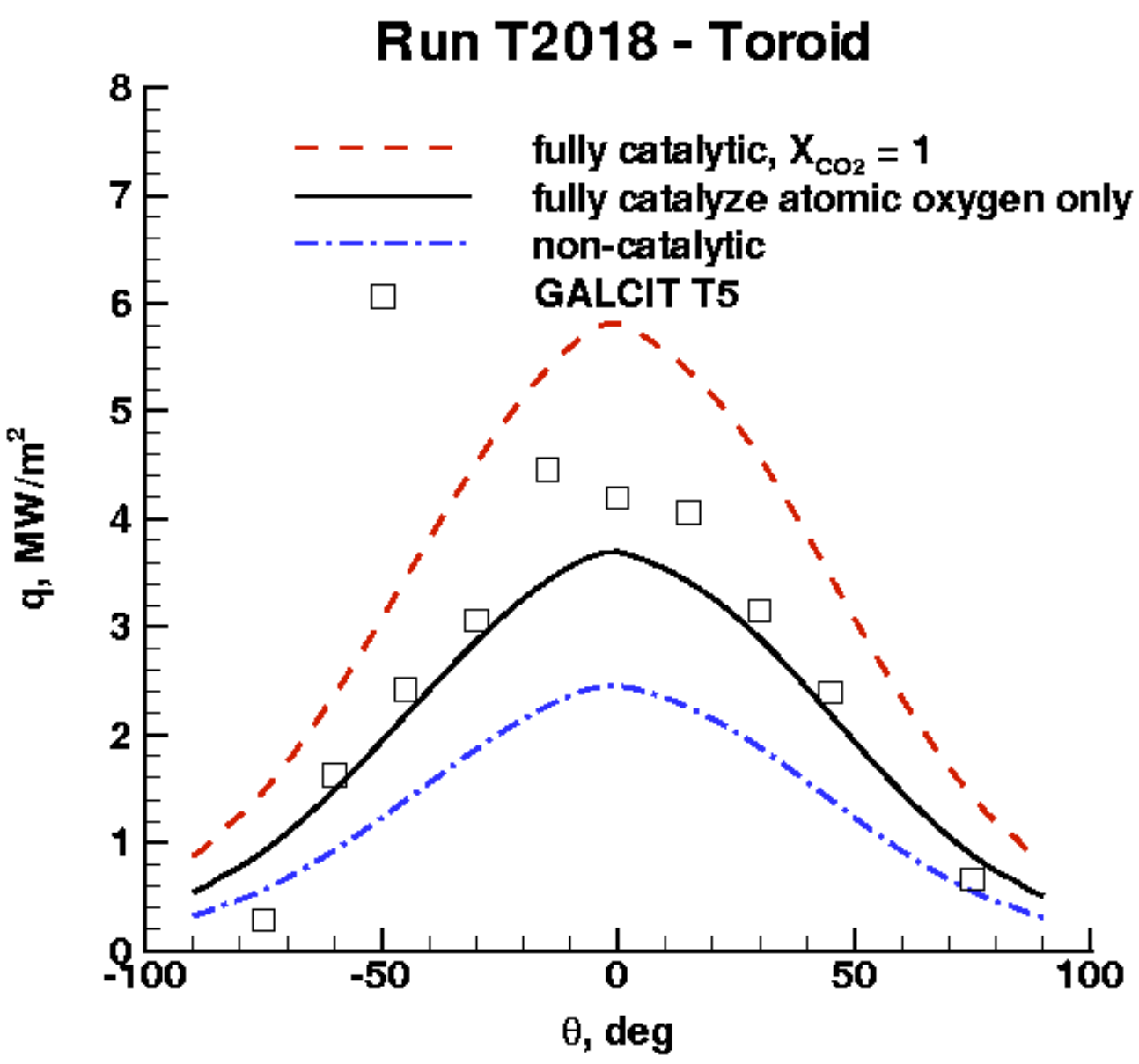

**Figure 11: Heat transfer over a toroidal ballute measured in T5 and computed [9]**

The plasma conditions surrounding the spacecraft (and the device) can be estimated using literature data, and if necessary, available tools for trajectory analysis and aerothermal calculations. Available experimental data will be considered, data from shock-tube tests are more relevant for chemical kinetics and radiation [25-26], however several test campaigns have been conducted in shock-tunnels for ballutes [22-23-27] and they have to be accounted for.

Disk and parachute-like configurations have also been simulated, but they put in evidence concerns about flow and aerodynamic stability. The same remark has been already reported for blunt body flows with cavities [28]. Other calculations have been performed for investigating the aerocapture of the MSRO aeroshell. They have been compared against NASA CF4 tunnel experiments.

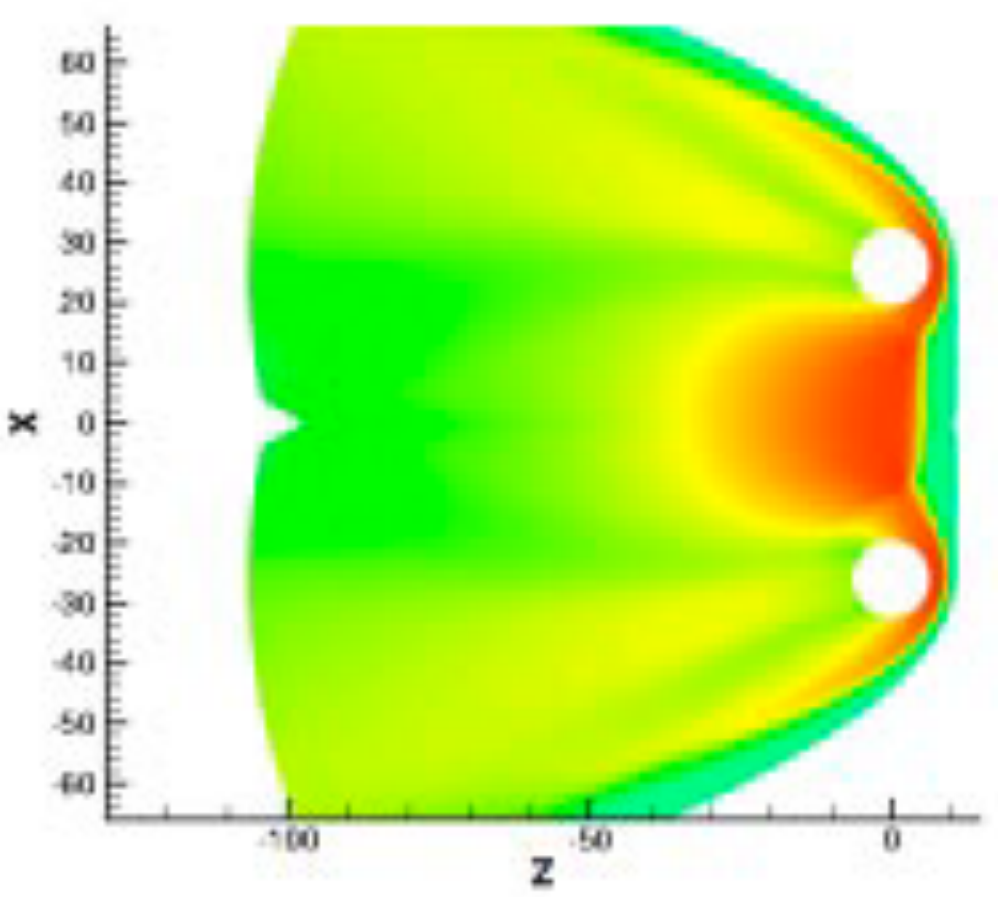

**Figure 12: Pressure contours for air (perfect gas, 5 km/s) at 2° angle of attack [8]**

The aerothermodynamics of spherical and toroidal towed ballutes have been considered in [8] with an emphasis of interactions between the spacecraft and the ballute. Continuum flow analysis was performed for towed and clamped ballute systems, in which the ballute can be quickly cut free when the requisite change in velocity is achieved. Both spherical and toroidal configurations have been computed for investigating the interactions between the ballute and the probe. Simulations show, as displayed in Figure 12, that for a toroid configuration, as the radius of the hole decreases, the flow through it will choke. Core flow is also influenced by the aerochemistry. For example for Titan calculations, the thermochemical non-equilibrium induces a smaller effective $\gamma$, promoting a shorter shock stand-off with shocks converging back to the axis, this enhances stronger interaction with a larger Mach disc in the core flow.

Ideally, interactions are avoided if the spacecraft wake (including bow shock) passes cleanly through the hole of the toroid (and towline wake effects can be ignored). Adverse interactions are encountered in the toroidal system if significant choking occurs in the core; however, these simulations [8] indicate that such adverse interactions can be avoided. The most significant adverse interaction appears to be a reverse flow back through the core of the spacecraft wake that may require increased thermal protection to the spacecraft base.

Interaction effects as a function of towline length were studied by Gnoffo & Anderson [8] for spherical ballute systems comprised of a 3 m radius spacecraft and a 35 m radius ballute. Separation distances from 30 to 200 m were included in the study. At these tow distances, the spherical ballute encounters a shock impingement that leads to unacceptably high localized heating and/or compromised drag effectiveness as the ballute encounters reduced dynamic pressure in the spacecraft wake.

| | Mars Micro Satellite | Titan Organics Explorer | Neptune Orbiter |
|---|---|---|---|
| $h_o$ (MJ/kg) | 15.1 | 36.6 | 361.8 |
| $U_{eqv}$ (m/s) | 5490 | 8550 | 26900 |
| $\rho'$ (kg/m$^3$) | $1.8 \times 10^{-5}$ | $4.7 \times 10^{-6}$ | $6.7 \times 10^{-8}$ |
| $Re'$ | 1400 | 1700 | 570 |
| $Re_\infty$ | 650 | 760 | 490 |

**Table 4: Relevant scaling parameters for the considered missions [22]**

| | Moderate Enthalpy 4.0 MJ/kg | Low Enthalpy 2.4 MJ/kg |
|---|---|---|
| Density (kg.m$^{-3}$) | 0.030 | 0.051 |
| Velocity (m.s$^{-1}$) | 2700 | 2100 |
| Pressure (kPa) | 3.1 | 3.0 |
| Temperature (K) | 347 | 200 |
| Mach Number | 7.5 | 7.1 |

**Table 5: T3 free-stream test conditions [27]**

In both the toroidal and spherical simulations [8], regions of high Knudsen number in the spacecraft wake were computed in which, the continuum Navier-Stokes simulations are compromised. Rarefied effects have been evaluated for a Neptune aerocapture by McRonald [29]. The mean free path of the free-stream near release of a typical ballute is about 1 m, so that the flow then is approximately continuum, while it will be free-molecular on the tubing and intermediate over the orbiter. In the entry case however, much of the drag occurs in developed continuum conditions, and a ring shaped ballute is indicated. The flow regime [30], with Knudsen number of 0.01 on the ballute and 0.1 on the orbiter, makes uncertain in what extent the orbiter wake reaches the ballute.

Finally, a very positive variation of the towed spherical ballute system employs a zero towline length (clamped system). In this design the spacecraft is contained fully within the low subsonic shock layer of the ballute and is nearly engulfed within the ballute boundary layer. Heating to the spacecraft forebody is reduced by a factor of two or greater (for the particular system considered here) and no adverse aeroheating is induced on the ballute.

Rasheed et al [22] have carried out series of experiments in the T5 Hypervelocity Shock Tunnel with for objective to determine the heat transfer rates to the ring of a toroidal ballute. Tests were performed with $CO_2$ with enthalpies ranging from 12 to 23 MJ/kg and reservoir pressure from 5 MPa to 30 MPa, nitrogen from 23 to 26 MJ/kg and 4 to 10 MPa, and in hydrogen from 27 up to 80 MJ/kg and 3 to 25 MPa, in order to support potential missions to Mars, Titan, and Neptune. The desirable conditions to support these missions are reported in Table 4. Collection of shadowgraphs and heating data was collected during the tests. Some of the shadowgraphs are shown in Figure 8 for $CO_2$ flows, and Figure 9 for $N_2$. Experimental values for heat-flux were in good agreement with theoretical values for $CO_2$ and $N_2$. The presence of unsteadiness was confirmed for some flow conditions, as highlighted in Figure 9. Such phenomenon was expected from the existing computational results of Gnoffo [9]. The tests confirmed that a toroidal ballute can avoid the unsteadiness encountered with simply connected ballute flows. This unsteadiness was resulting from the shock-shock interaction between the ballute and the spacecraft.

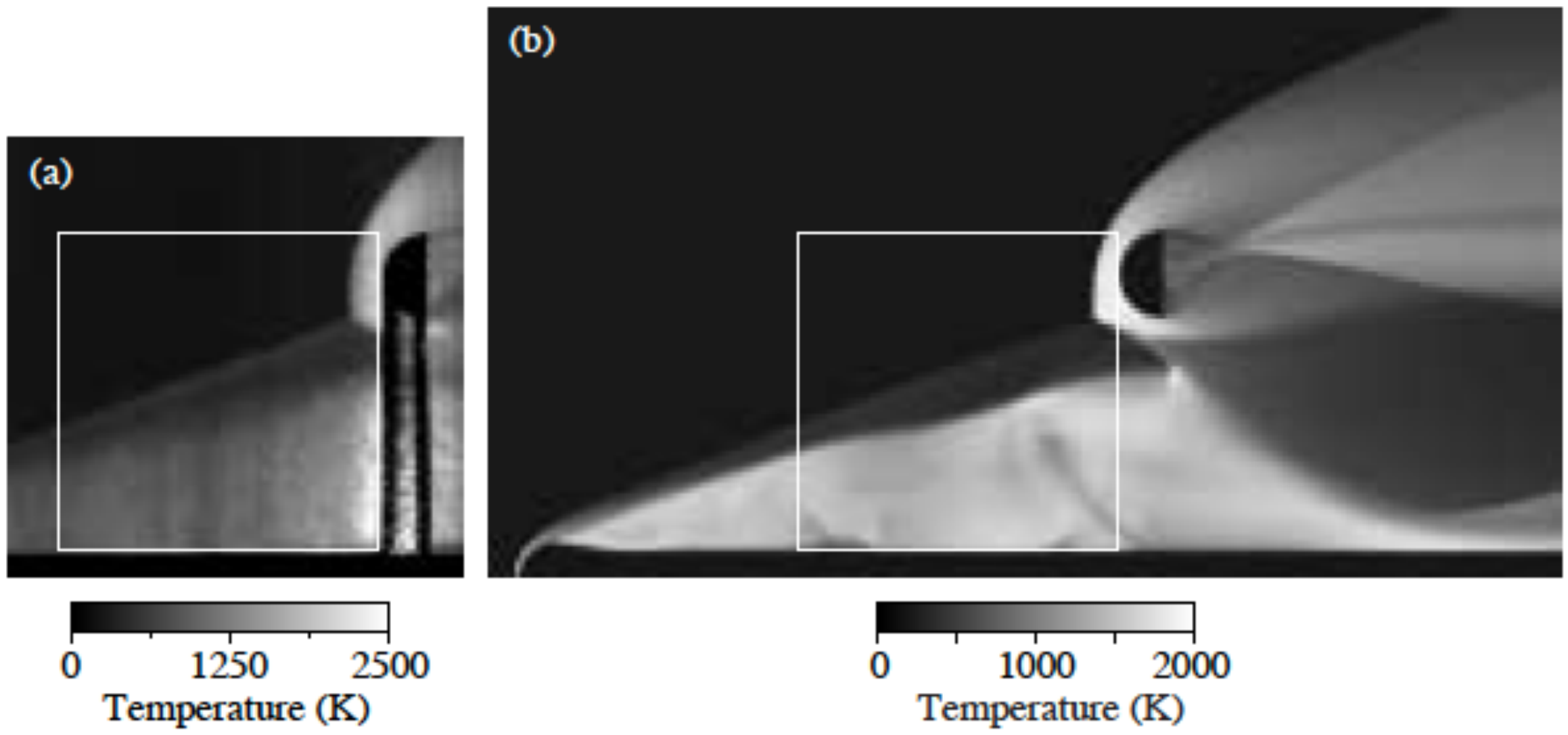

**Figure 13: Comparison between temperature measurement (left) and CFD calculation. Rectangle indicates common region [27]**

| Condition | Gas | Enthalpy, MJ/kg | Binary scaling $\rho_e r$, kg/m$^2$ | Mach number |
|---|---|---|---|---|
| A | $CO_2$ | 17–18 | $(1.2–1.6) \times 10^{-4}$ | 7.7–8.3 |
| B | $CO_2$ | 48–52 | $(8–15) \times 10^{-5}$ | 9.3–10.3 |
| C | $N_2$ | 17–19 | $(6–8) \times 10^{-5}$ | 7.2–8.1 |
| D | $N_2$ | 52–58 | $(6–11) \times 10^{-5}$ | 8.1–9.0 |
| E | $H_2$/Ne[a] | 79–82 | $(2–3) \times 10^{-5}$ | 8.2–8.7 |

[a]Mixture for condition E was 15% $H_2$, 85% Ne.

**Table 6: X2 free-stream conditions [23]**

Lourel et al [27] have carried out experiments in T3 shock-tunnel to investigate flows around a toroidal ballute and a towed ballute-spacecraft system. Fluorescence imaging and thermometry measurements were used to visualize and obtain temperature maps of the flow-fields. Tests were performed with $N_2$, test conditions are reported in Table 5. The separation length between the ballute and the spacecraft was varied and both steady and unsteady flow-fields were observed. Heat transfer measurements, PLIF images and comparisons against CFD results are presented in Figure 13. This figure reveals the complex intricate flow patterns around probe and ballute. This study confirms the conclusions of the survey of Hall [11] about the considerable amount of work remaining to verify the ballute concept.

Another test campaign, on toroidal ballute, has been conducted in X2 expansion tube [23] at moderate enthalpy conditions (15-20 MJ/kg) and high enthalpies (> 50 MJ/kg). At moderate enthalpies tests were performed in $CO_2$ and $N_2$, at higher enthalpies a mixture of $H_2$-Ne was also investigated, the test matrix is reported in Table 6. Heat flux measurements have been performed as well as imaging with holographic interferometry. Series of tests were done with blocked and unblocked toroids. Results show, as highlighted in Figure 14, that the blocked configuration is more unstable due to the interaction between the blocked toroid and bow shock. Generally, measure heat-transfer values were in good agreement with computed ones.

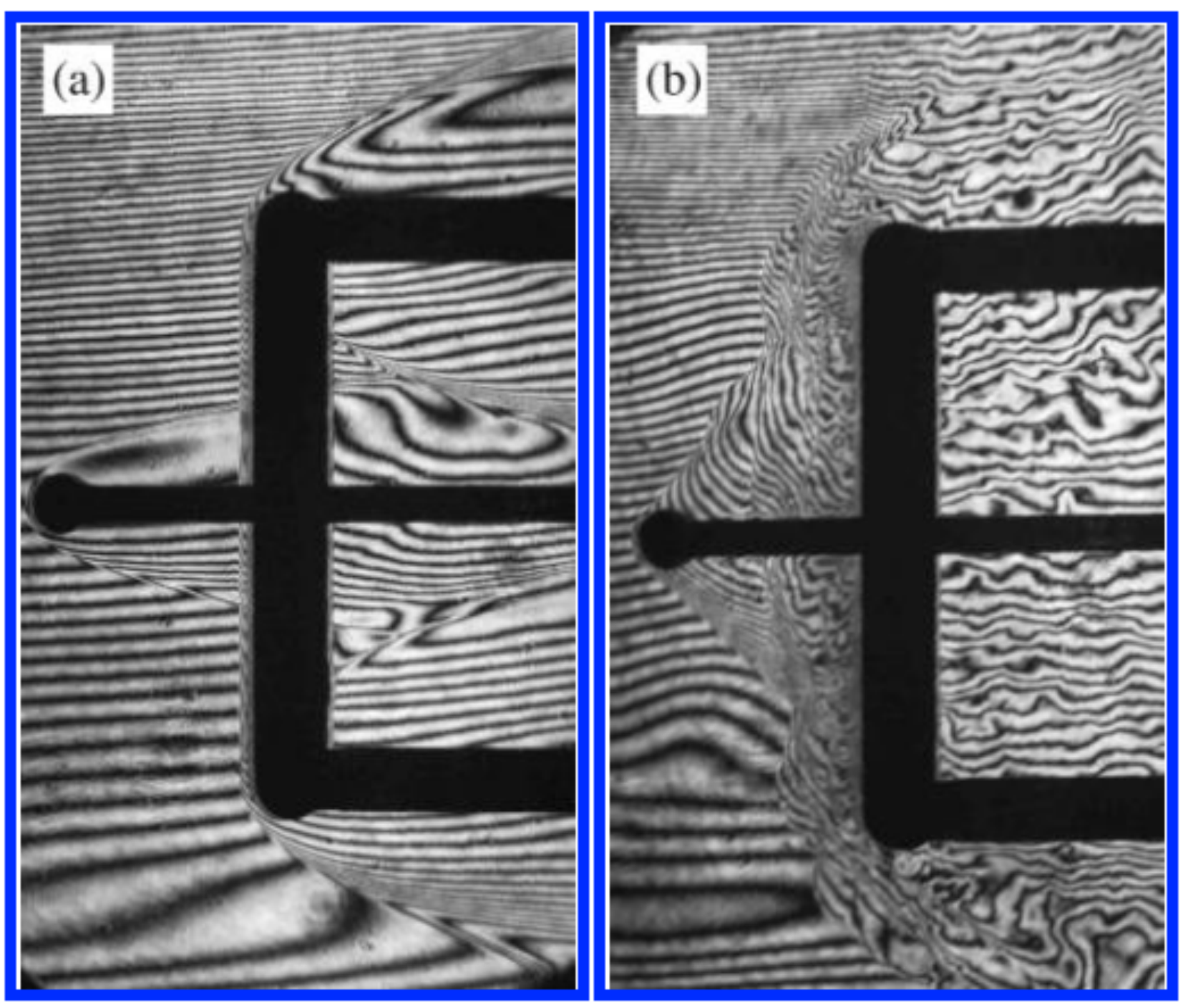

**Figure 14: Flow-fields obtained with holographic interferometry for Case B of Table 6[23] (left unblocked model, right blocked)**

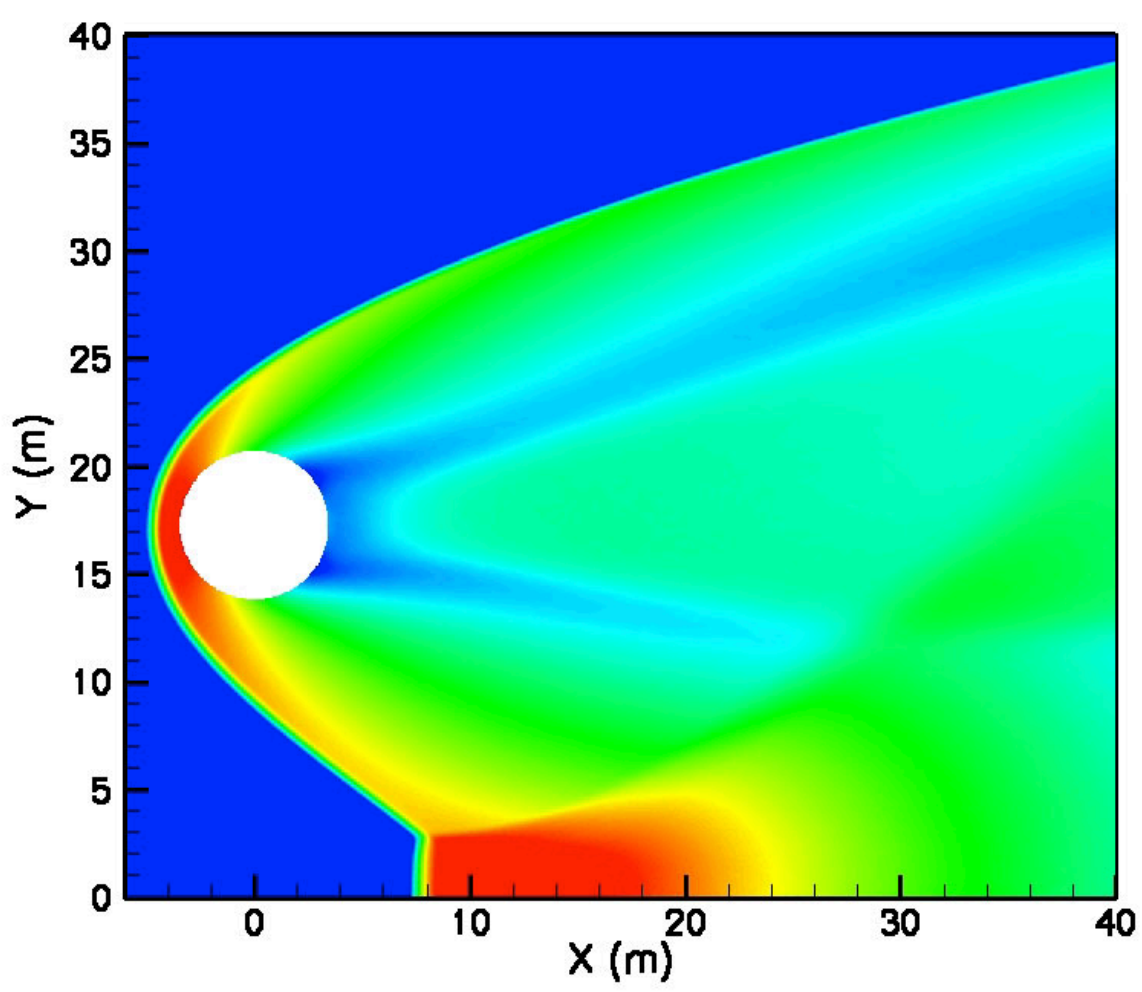

**Figure 15: Pressure contours predicted using DSMC near peaking conditions for a Titan aerocapture (5.52 km/s, 168 K) [7]**

Calculations with DSMC and continuum approaches for a Titan aerocapture have been reported by Miller et al [7]. Figure 15 shows the contours obtained for the pressure for a Titan aerocapture at peaking conditions. The choked flows located in the center of the toroid is clearly defined. Series of calculations were carried out for comparing the predictions of the drag. The assessment highlights the validity of using bridging functions. The calculations allow to modify them in order to optimize the numerical predictions.

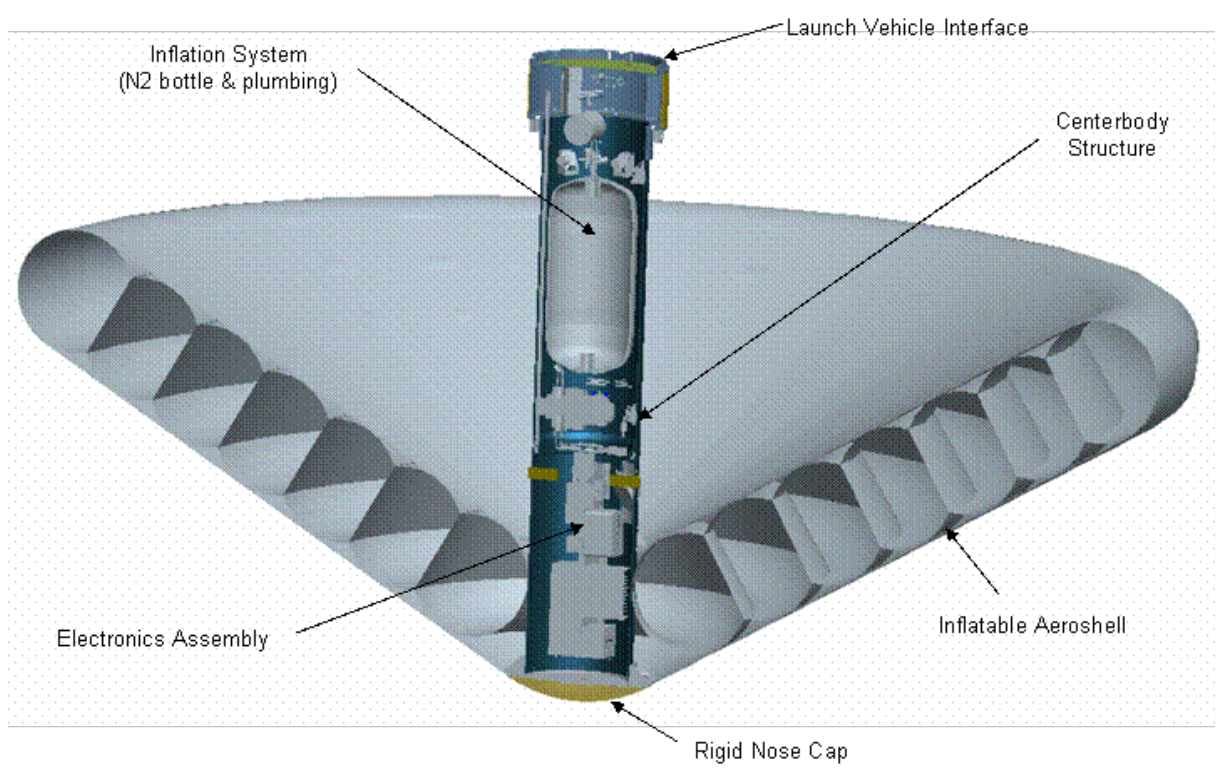

**Figure 16: Scheme of IRVE capsule [31]**

## 4.2 Inflatables

Concerning inflatable devices, so far IRDT [4-5] and IRVE were the only flight tests involving an atmospheric entry of such devices. IRDT capsules, shown in Figure 2, were developed by the Babakin Space Centre under ESA funding. On her side, NASA [31] has developed, later on, the IRVE (Inflatable Re-entry Vehicle Experiment) vehicle shown in Figure 16. Little elements have been found in the literature related to NASA experiments. Concerning IRDT, a test campaign was performed in shock-tube at HTG Göttingen (reported in [4]), but little details have been published on it. No information is available of potential test campaigns carried out in Russia. Concerning CFD analysis, numerical simulations were performed for supporting the project preparation [4] and the post-flight analysis [5]. For the project support, the aerothermal effort was focusing on the assessment of transition to turbulence over the capsule, post-flight calculations were dedicated to evaluate to ionization near the front-shield of the IRDT, where the antenna was embedded, to analyze the blackout duration observed during the flight. Inflatable aspects were also investigated for the YES2 project developed by ESA. FEM calculations [32] included attempt to simulate the fluid-structure interaction [33] were performed.

An overview on inflatable structures has been published by Veldman and Vermeeren [34], however the paper is limited in the way that pros and cons are not identified, and most of the critical aspects such as aerothermodynamics as well as structure and material issues are not identified. A survey of the critical aspects and key technologies for inflatables has been performed recently by Overend et al [35] and Underwood et al [36]. A system study was carried out for the potential missions targeted by ESA, and current capabilities for test campaigns in Europe were reviewed. However, the heritage of the IRDT project including lessons learnt was not reviewed. A roadmap for future activities was proposed. In the frame of this study an experimental campaign was conducted in LBK at DLR, for testing insulating material (see Figure 17).

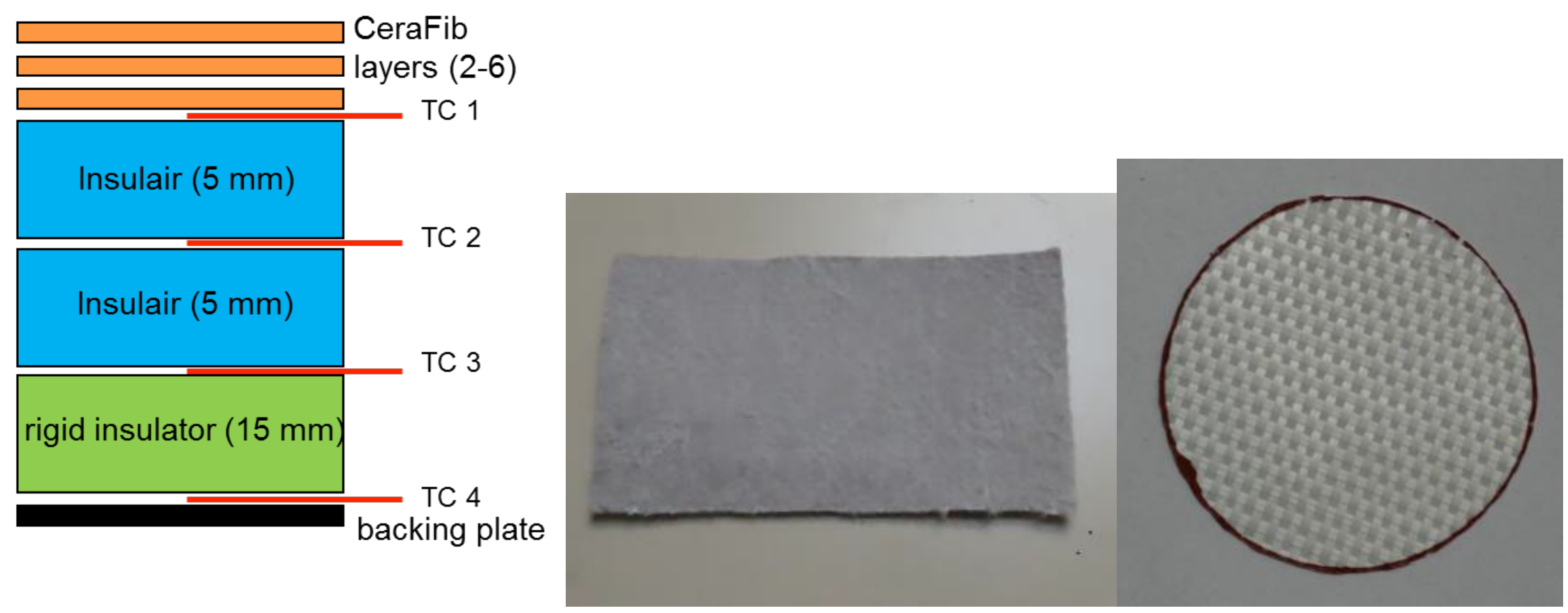

**Figure 17: Insulating TPS tested at DLR: from left to right, flexible TPS, Insulair NP650, and Cerafib Tex99 [35]**

The interest of inflatables for aerobraking and aerocapture has never been studied in details, indeed, this kind of structure has to remain deployed around the spacecraft, as a consequence, it fits better for atmospheric entry. Their interest for Mars and Venus entries has been investigated by Dutta et al [37]. *The advantages over rigid aeroshells are more pronounced when the diameters of the deployable bodies are increased, leading to a larger wetted area, lower areal density and higher payload mass fraction. However, the shallow angle of entry opens the possibilities of skip-out during entry; thus, a skip-out margin has to be defined based on deviations from the nominal due to atmospheric perturbations or interplanetary trajectory delivery errors. Additionally, in the Venusian atmosphere, non-equilibrium radiation could be enhanced, especially since the deployables will have large diameters (and nose radii). Finally, there are still some unresolved issues with both static and dynamic stability of these vehicles, especially since the flexible body will interact with pressure loadings.*

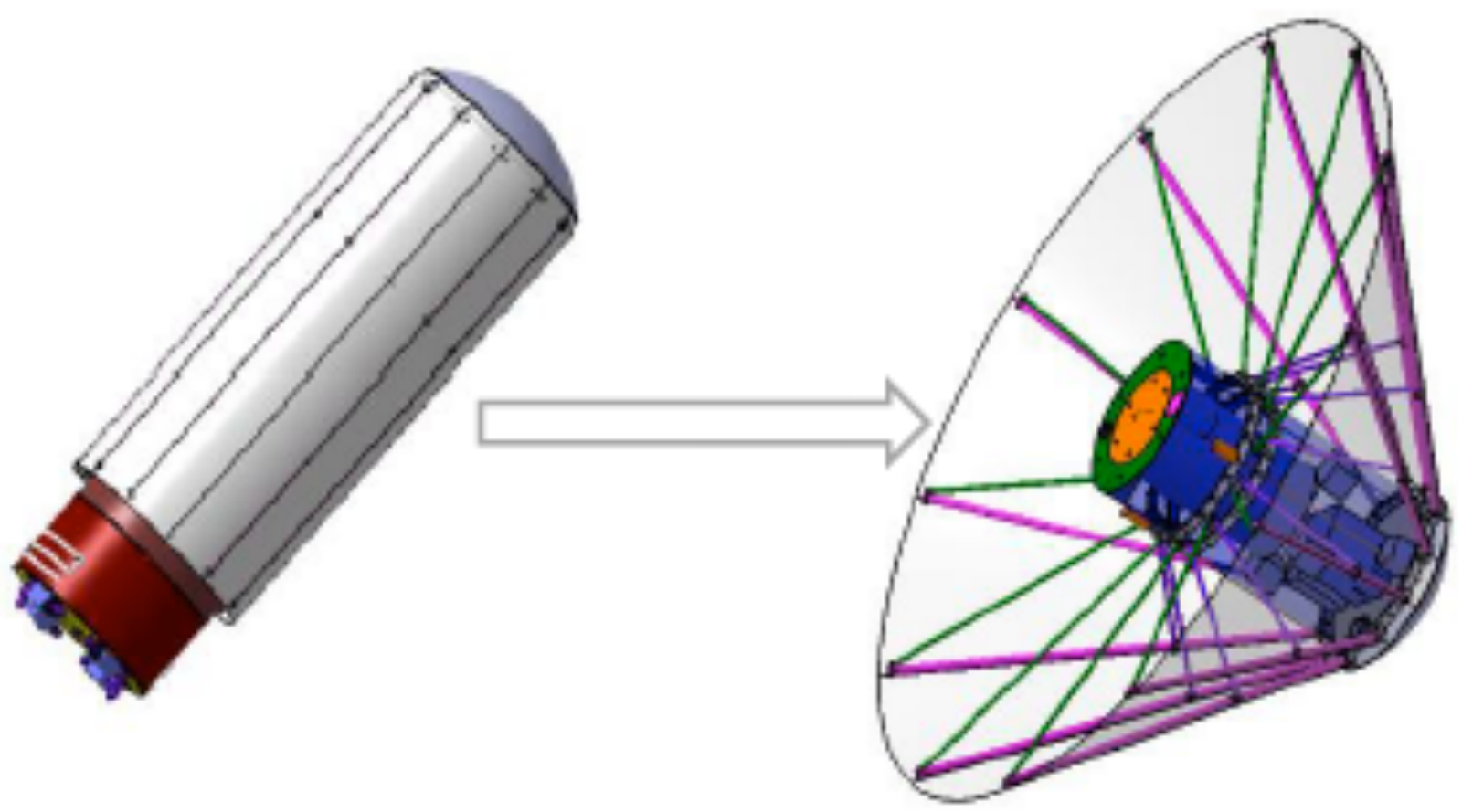

**Figure 18: IRENE deployable capsule [39]**

### 4.3 Deployable sails

In the recent years ESA and ASI have fostered the development of re-entry capsule with deployable TPS. The Mini-IRENE project [38-39] consists in a heat-shield with rigid and deployable parts as shown in Figure 18. The objective is to perform a sub-orbital flight test using a Maxus sounding rocket. TPS qualification has been performed in SCIROCCO PWT [39], deployment was tested in vacuum conditions. FEM calculations have been performed as highlighted in Figure 19. Details on the different subsystems are given in [38]. However, the current development of deployable sails is fostered by their use as deorbiting device for satellites at their end-of-life. .

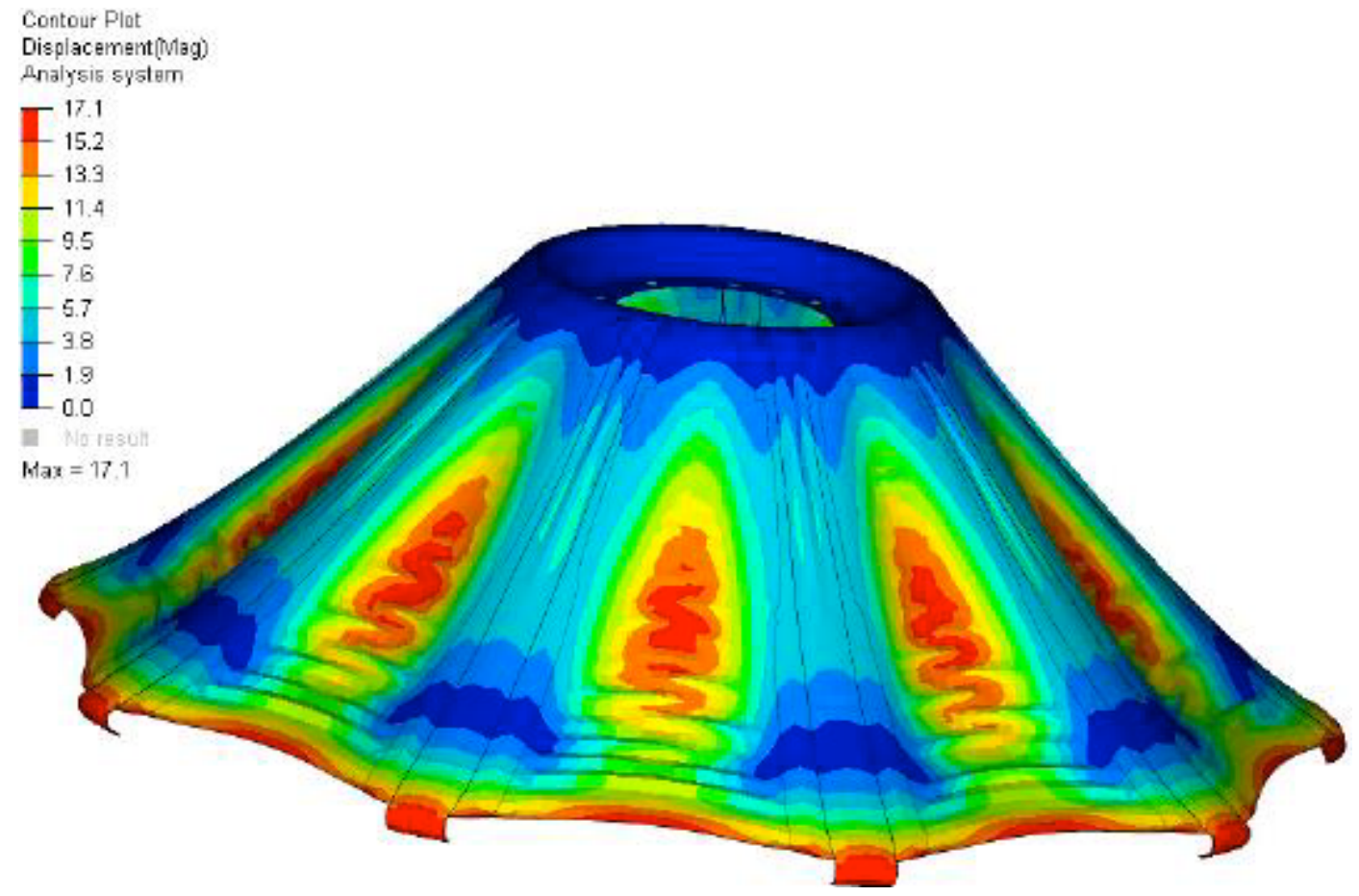

**Figure 19: FEM calculation of IRENE capsule [38]**

## 5 Conclusions

A survey of the current state-of-the-art for inflatable and deployable technologies has been performed in the frame of this internal study. The interest of these different options for aerobraking and aerocapture is assessed. Ballute technology has been largely studied in the United States, while little elements can be found in Europe. Deployable sails can be an option particularly for aerobraking, they are easier to be enforced since they are less complex than inflatable devices. It has to be noted that European activities have some limitations. Due to the lack of timelessness of the effort, there is a difficulty to capitalize on the experience and lessons learnt. Concerning aerobraking, sails and towed ballutes seem to be the best options, while for aerocapture ballutes fits apparently better, particularly if a toroidal shape is selected. The test campaigns for ballutes performed in USA and Australia have been considered and their main outcomes included in the current effort.